# Contrasting the relative performance of RF photonic transversal signal processors based on microcombs using discrete components versus integrated devices

David J. Moss

Optical Sciences Center, Swinburne University of Technology, Hawthorn, VIC 3122, Australia

e-mail: dmoss@swin.edu.au.

**Abstract**

**RF photonic transversal signal processors, which combine reconfigurable electrical digital signal processing and high-bandwidth photonic processing, provide a powerful solution for achieving adaptive high-speed information processing. Recent progress in optical microcomb technology provides compelling multi-wavelength sources with compact footprint, yielding a variety of microcomb-based RF photonic transversal signal processors implemented by either discrete or integrated components. Although operating based on the same principle, processors in these two forms exhibit distinct performance. This letter presents a comparative investigation into their performance. First, we compare the performance of state-of-the-art processors, focusing on the processing accuracy. Next, we analyze various factors that contribute to the performance differences, including tap number and imperfect response of experimental components. Finally, we discuss the potential for future improvement. These results provide a comprehensive comparison of microcomb-based RF photonic transversal signal processors implemented using discrete and integrated components and provide insights for their future development.**

I. INTRODUCTION

Driven by the exponential growth of data capacity, there has been a rapid increase in the demand for high-speed information processing. Radio-frequency (RF) photonics, which utilize photonic hardware and technologies to process high-bandwidth RF signals, provides speed advantages over electrical signal processing with intrinsic bandwidth limitations [1]. Among the various schemes to implement RF photonic processors, RF photonic transversal signal processors have garnered significant interest due to their exceptional reconfigurability, allowing for the realization of diverse processing functions without the need for changing any hardware [2-36].

In RF photonic transversal signal processors, a number of wavelength channels are needed to facilitate high reconfigurability and ensure high processing accuracy. In addition, a wide channel spacing between these wavelength channels is necessary to ensure a large operational bandwidth. Recent advances in optical microcomb technology provide a competitive solution to meet these requirements by generating a substantial number of widely spaced wavelengths from a micro-scale resonator, together with added benefits of significantly reduced device footprint, power consumption, and complexity [37]. In contrast, conventional multi-wavelength sources, such as discrete laser arrays [38], fiber Bragg grating arrays [39], laser frequency combs generated by electro-optic (EO) modulation [40], and mode-locked fiber lasers [41] suffer from limitations in one form or another, such as a limited number of wavelength channels and insufficient channel spacings.

Early implementations of microcomb-based RF photonic transversal signal processors simply replaced conventional multi-wavelength sources with optical microcombs while retaining all other components as discrete devices [1, 2-36]. Although this already yields significant benefits, there is much more to be gained by increasing the level of integration for the entire processing system, particularly with respect to the system size, power consumption, and cost. Recently, several processors comprised entirely of integrated components have also been demonstrated [42, 43]. Despite based on the same operation principle, the processors implemented by discrete and integrated components present different processing performance.

In this paper, we provide a comparative study of the performance of microcomb-based RF photonic transversal signal processors implemented by discrete and integrated components. First, we compare the performance of state-of-the-art processors, focusing on the processing accuracy. Next, we conduct analysis of multiple factors that induce the performance differences, including tap number and imperfect response of experimental components. Finally, we discuss the potential for future development. These results provide a comprehensive comparison and valuable perspectives for these processors with high reconfigurability for diverse signal processing applications.

## II. MICROCOMB-BASED RF PHOTONIC TRANSVERSAL SIGNAL PROCESSORS

RF transversal signal processors are implemented based on the transversal filter structure in digital signal processing, which features a finite impulse response and has found applications in a wide range of signal processing functions [1]. Implementing these processors by using RF photonic technology can yield significantly higher processing bandwidth compared to their electronic counterparts [1], and the use of optical microcombs provides a powerful multiwavelength source that is critical for the RF photonic system. **Fig. 1(a)** illustrates the operation principle of a microcomb-based RF photonic transversal signal processor. The processor employs an optical microcomb as a multiwavelength source, which simultaneously generates numerous wavelength channels as discrete taps. Input RF signal is modulated onto each wavelength channel via EO modulation, producing multiple RF replicas. Next, optical spectral shaping is applied to weight these modulated replicas, and time delay is introduced between adjacent wavelength channels. Finally, the weighted and delayed RF replicas are added together through photodetection to generate the final RF output of the processor. After going through the processing flow in **Fig. 1(a)**, the output RF signal $s(t)$ can be given by [1].

$$s(t) = \sum_{n=0}^{M-1} a_n f(t - n\Delta T), \tag{1}$$

where $f(t)$ is the input RF signal, $M$ is the tap number, $a_n$ ($n$ = 0, 1, 2, …, $M$-1) is the tap weight of the $n^{th}$ tap, and $\Delta T$ is the time delay between adjacent wavelength channels. Therefore, the system's impulse response can be expressed as [1].

$$h(t) = \sum_{n=0}^{M-1} a_n \delta(t - n\Delta T), \tag{2}$$

After Fourier transformation from **Eq. (2)**, the spectral transfer function of the processor can be described as

$$H(\omega) = \text{FT}\,[h(t)] = \sum_{n=0}^{M-1} a_n e^{-j\omega n \Delta T}, \tag{3}$$

According to **Eqs. (1) – (3)**, different processing functions can be realized by appropriately setting the tap coefficients $a_n$ ($n$ = 0, 1, 2, …, $M$-1) without changing the hardware, which allows for a high reconfigurability for the processor.

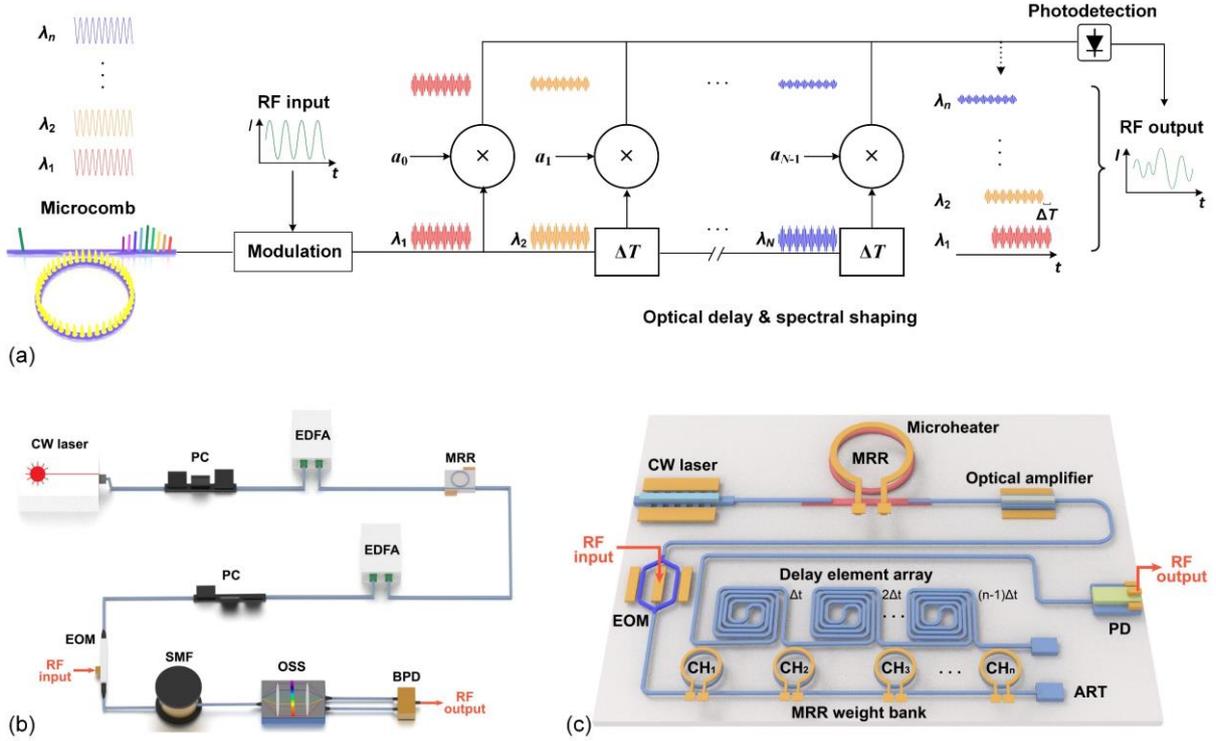

**Fig. 1.** (a) Schematic illustration of the operation principle of a microcomb-based RF photonic transversal signal processor. (b) Schematic of a microcomb-based RF photonic transversal signal processor implemented by discrete components. (c) Schematic of an on-chip microcomb-based RF photonic transversal signal processor implemented by integrated components. EOM: electro-optic modulator. RF: radio frequency. PD: photodetector. CW laser: continuous-wave laser. EDFA: erbium-doped fibre amplifier. PC: polarization controller. MRR: microring resonator. SMF: single-mode fibre. OSS: optical spectral shaper. BPD: balanced photodetector. BPD: balanced photodetector. ART: anti-reflection termination.

Based on the operation principle in **Fig. 1(a)**, microcomb-based RF photonic transversal signal processors can be practically implemented into two forms. The first form is illustrated in **Fig. 1(b)**, where all the components are discrete devices except for an integrated microring resonator (MRR) used to generate optical microcombs. The second is comprised entirely of integrated components, as shown in **Fig. 1(c)**. To simplify our discussion, we will refer to the processors implemented in these two forms as discrete and integrated processors. Early microcomb-based RF photonic transversal signal processors were in the form of discrete processors [1, 2], while more recently, some integrated processors have been demonstrated [42, 43]. Although the operation principle remains consistent across these two forms, their different components result in distinct performance. In the following, we compare their performance in **Section III** and discuss their potential for improvement in **Section IV**.

III. PERFORMANCE COMPARISON OF DISCRETE AND INTEGRATED PROCESSORS

In this section, we compare the performance of discrete and integrated processors shown in **Figs. 1(b)** and **(c)**, respectively. Although the size, weight, and power consumption (SWaP) of integrated processors are greatly reduced compared with the discrete processors, the state-of-the-art integrated processors suffer from limited tap numbers due to the restrictions imposed by the integrated components. Currently, integrated processors with only 8 [42] and 12 taps [43] have been demonstrated, whereas discrete processors have been implemented with up to 80 taps [1, 44]. The difference in the tap numbers results in a difference in the processing accuracy [44]. In addition, the imperfect response of experimental components also induces processing errors. These mainly include noise of microcombs, chirp of the EOM, errors in the delay element, and errors of the spectral shaping module [44].

**Table I** summarizes parameters of the components in the three processors that we investigate, including a discrete processor (Processor 1) and two integrated processors (Processors 2 and 3). There are two integrated processors: one with the same tap number as that in Ref. [42] and the other with an increased tap number to demonstrate the potential for improvement. To characterize the errors induced by imperfect experimental components, optical signal-to-noise ratios (OSNRs), chirp parameters ($\alpha$), error of the delay element ($t_v$), and random tap coefficient errors (RTCEs) are introduced. These parameters are all set based on the real processors in Refs. [43-47]. For comparison, we assume that the three processors employ the same microcomb with a comb spacing of 0.4 nm (~50 GHz). We also assume that the time delay between adjacent taps in **Eq. (3)** is $\Delta T = 33.4$ ps to ensure the same operation bandwidth.

**TABLE I. COMPARISON OF COMPONENTS' PARAMETERS IN DISCRETE AND INTEGRATED PROCESSORS**

| Discrete processor | No. | Tap No. | OSNR of microcombs | Chirp parameter of the EOM | Errors of the delay element | RTCE of the spectral shaping module |
|---|---|---|---|---|---|---|
| | 1 | $M = 80$ [44] | $OSNR$ : 20 dB [44] | $\alpha$ : 0.1 [47] | $t_v$ : 4% [44] | $RTCE$ : 5% [44] |
| Integrated processors | No. | Tap No. | OSNR of microcombs | Chirp parameter of the EOM | Error of the delay element | RTCE of the spectral shaping module |
| | 2 | $M = 8$ [42] | $OSNR$ : 20 dB [44] | $\alpha$ : 0.8 [45] | $t_v$ : 3% [43] | $RTCE$ : 9% [46] |
| | No. | Tap No. | OSNR of microcombs | Chirp parameter of the EOM | Error of the delay element | RTCE of the spectral shaping module |
| | 3 | $M = 20$ | $OSNR$ : 20 dB [44] | $\alpha$ : 0.8 [45] | $t_v$ : 3% [43] | $RTCE$ : 9% [46] |

Compared with discrete EOM, the integrated EOM has a relatively high chirp parameter in **Table I**, mainly because achieving an accurate bias point and precise electrode placement is more challenging for integrated devices [47]. The lower accuracy for the delay element in the discrete processor is mainly caused by the high-order dispersion of the dispersive medium (e.g., optical fibre), which results in non-uniform time delays between adjacent wavelength channels [44]. In contrast, in integrated processors time delay is introduced by integrated optical delay lines (e.g., Si spiral waveguides), which exhibit a higher accuracy owing to the precise control over the amount of delay achieved by designing specific length and refractive index profile [48]. The accuracy differences between the integrated and discrete spectral shaping modules are more noticeable. Although the integrated spectral shaping modules have much lower tap numbers, they have lower spectral shaping accuracy compared to their discrete counterparts. This is due to the fact that commercial discrete waveshapers based on mature liquid crystal on silicon (LCoS) technology offer much better accuracy for amplitude and phase control, as well as inter-channel synchronization [49].

In our following analysis, three typical signal processing functions including first-order differentiation (DIF), integration (INT), and Hilbert transform (HT) are taken as examples to compare the accuracy of discrete and integrated processors. The tap numbers required to achieve these processing functions are designed based on our previous work in Refs. [1]. To quantify the comparison of processing accuracy, the root mean square error (RMSE) is introduced to compare the deviation between the processors' outputs and the ideal results, which is expressed as

$$\text{RMSE} = \sqrt{\sum_{i=1}^{k} \frac{(Y_i - y_i)^2}{k}} \qquad (4)$$

where $k$ is the number of sampled points, $Y_1, Y_2, \ldots, Y_n$ are the values of the ideal result, and $y_1, y_2, \ldots, y_n$ are the values of the output of the processors.

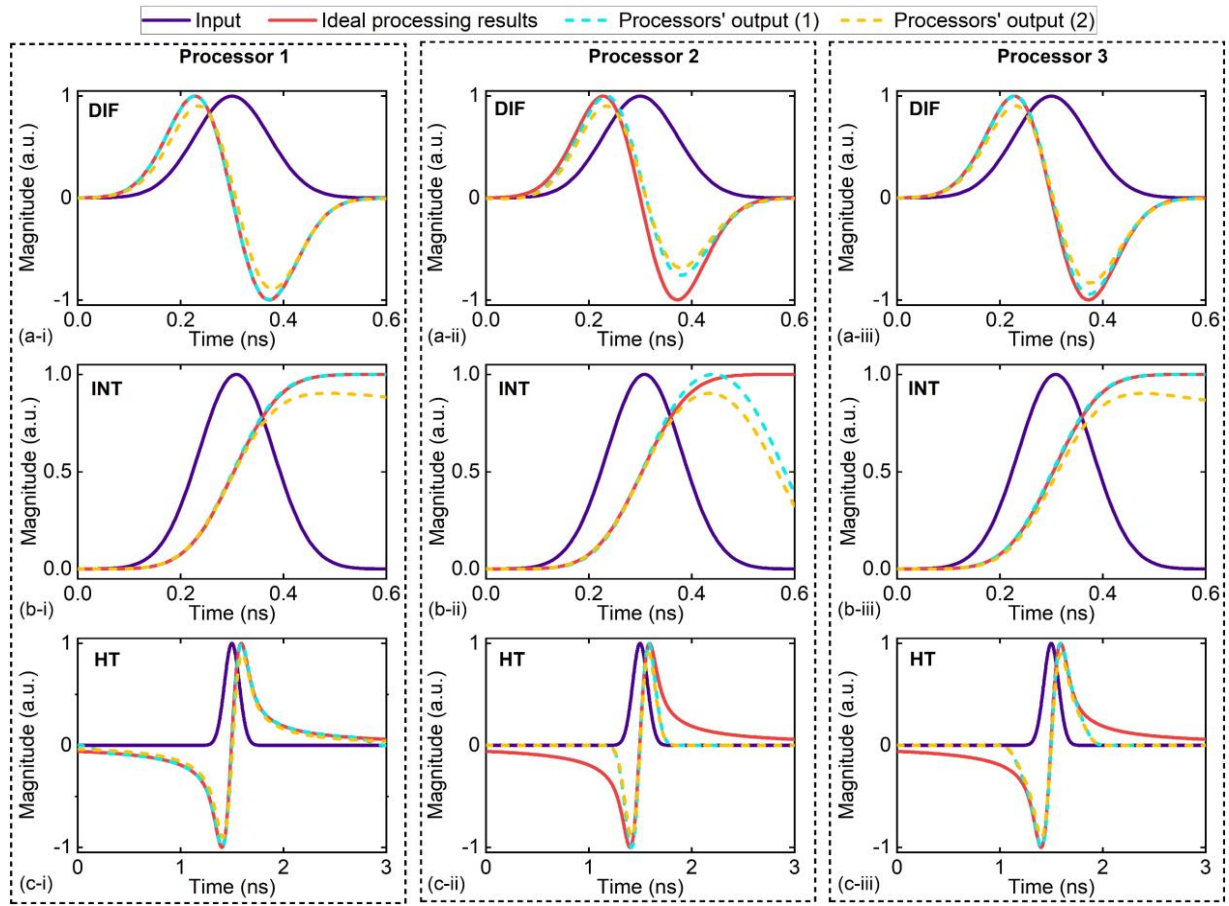

**Fig. 2.** Temporal waveform of Gaussian input pulse and output waveforms from Processors 1 – 3 that perform (a) differentiation (DIF), (b) integration (INT), and (c) Hilbert transform (HT). The processors' outputs with errors induced by (1) only limited tap numbers and (2) both limited tap numbers and experimental errors are shown, together with the ideal processing result for comparison.

**Figs. 2(a) – (c)** show the outputs of Processors 1 – 3 in **Table I** that perform DIF, INT, and HT, respectively. The input RF signal is a Gaussian pulse with a full-width-at-half-maximum (FWHM) of ~0.17 ns. Here we show the processors' outputs with errors induced by (1) only limited tap numbers and (2) both limited tap numbers and experimental errors. The ideal processing results are also shown for comparison. Deviations between the processors' outputs and the ideal results are observed for all three functions, and the deviations become more significant when taking into account the experimental errors.

**Fig. 3** compares RMSEs of the processors in **Fig. 2**. The higher processing accuracy of the discrete processor, compared to the integrated processors, is reflected by the lower RMSEs of Processor 1 for all three processing functions. In addition, the RMSEs of Processor 3 are lower compared to Processor 2, which indicates a higher processing accuracy achieved by increasing the tap number. According to the results in **Fig. 3**, the primary factor that contributes to the degradation of accuracy for integrated processors is the limited tap number. Whereas for discrete processors with a sufficiently large tap number, the processing inaccuracy is mainly induced by the imperfect response of experimental components. We also note that the differences in RMSEs among Processors 1 – 3 are more prominent for the INT than the other two processing functions, indicating a higher requirement for a greater number of taps to improve the processing accuracy of INT. In addition, experimental errors have a substantial impact on the RMSEs of DIF, whereas their impact on HT is very small.

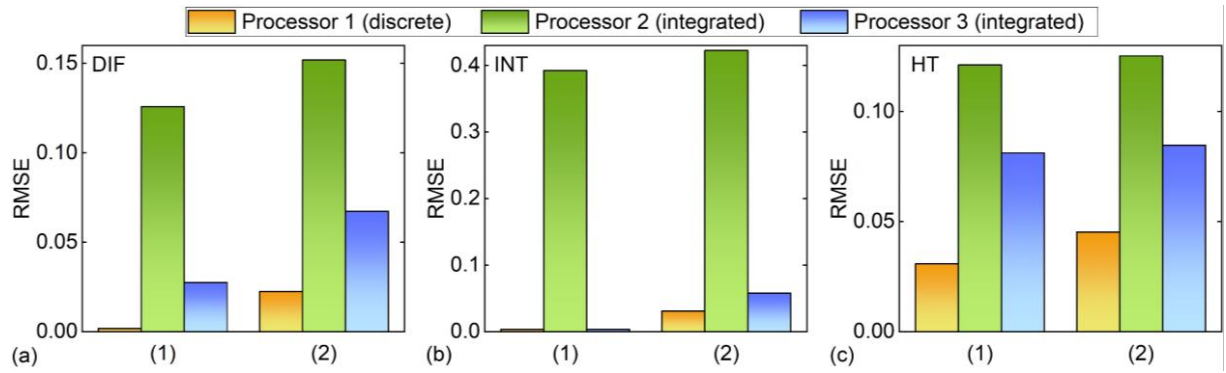

**Fig. 3.** Comparison of root mean square errors (RMSEs) of Processors 1 – 3 that perform (a) DIF, (b) INT, and (c) HT. The RMSEs of processors' outputs with errors induced by (1) only limited tap numbers and (2) both limited tap numbers and experimental errors are shown.

IV. POTENTIAL FOR IMPROVEMENT

In this section, we discuss the potential for improvement for both the discrete and integrated processors. In **Fig. 4(a)**, we quantitively analyze the influence of tap number $M$ on the processing accuracy, where the parameters of the input signal and the processor components are kept the same as those in **Fig. 2**. Similar to that in **Figs. 2** and **3**, we show the results with errors induced by (1) only limited tap numbers and (2) both limited tap numbers and experimental errors. It is evident that when assuming no experimental errors, both discrete and integrated processors exhibit the same RMSE values at the same $M$, as they possess identical comb spacing and time delay.

When experimental errors are considered, the RMSEs no longer exhibit a monotonic decrease with the tap number *M* as observed when assuming no experimental errors. This can be attributed to the accumulated experimental errors as *M* increases, such as increased shaping errors in both discrete and integrated processors, as well as the increased errors of time delay induced by higher-order dispersion in discrete processors [44].

When considering experimental errors, DIF, INT, and HT require a tap number of 20, 20, and 80 to achieve a RMSE of ~0.05, respectively. Although this can be easily achieved by the discrete processor, it is challenging for the state-of-the-art integrated processors due to the significantly increased complexity and degraded processing accuracy for $M \geq 20$. The increased complexity results from the increased numbers of MRRs, micro-heaters, and spiral waveguides in **Fig. 1(c)**. Although integrated processors have the advantage of monotonically integrating a large number of these building blocks, achieving their precise tuning and control can be challenging, especially when dealing with a large number of taps. On the other hand, fabrication errors, additional loss, and thermal drifts in these building blocks degrade the cooperative operation of different wavelength channels, and the processing errors resulting from these factors increase super linearly with the tap number.

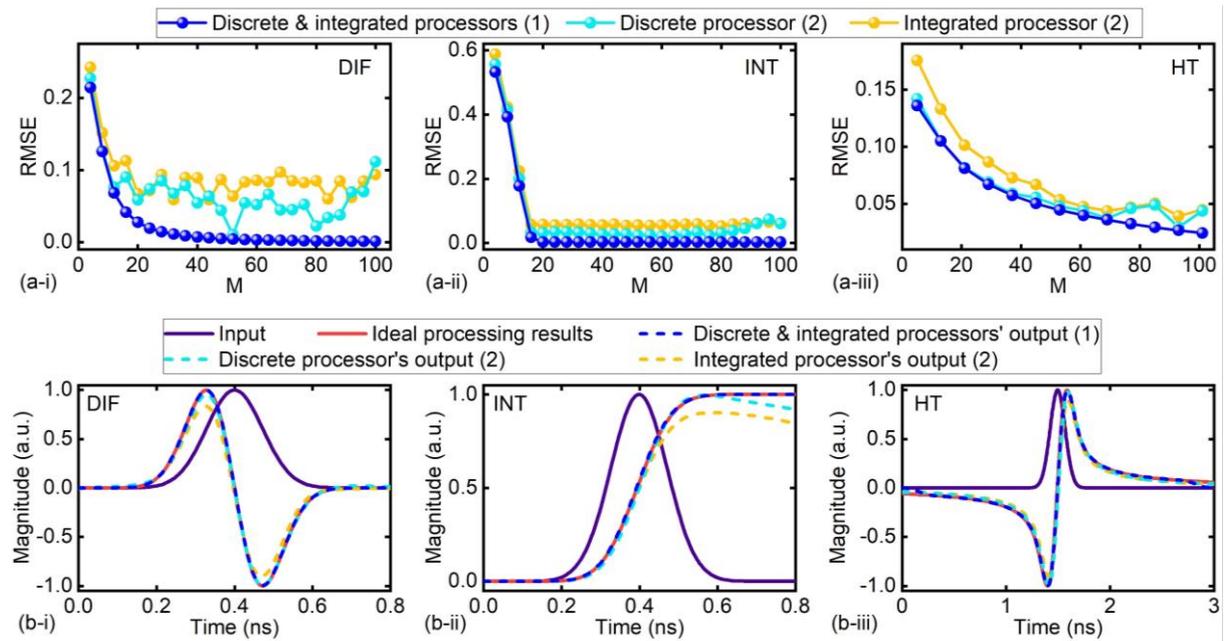

**Fig. 4.** Influence of the tap number and experimental errors on processing accuracy of the discrete and integrated processors. (a) RMSEs of processors that perform (i) DIF, (ii) INT, and (iii) HT as a function of tap number *M*. The RMSEs of processors' outputs with errors induced by (1) only limited tap numbers and (2) both limited tap numbers and experimental errors are shown. (b) Temporal waveform of Gaussian input pulse and output waveforms of discrete and integrated processors that perform (i) DIF, (ii) INT, and (iii) HT. The processors' outputs with errors induced by (1) limited tap number and (2) both the limited tap number and experimental errors are shown, together with the ideal processing result for comparison.

**Fig. 4(b)** compares the output waveforms of both discrete and integrated processors with the same tap number $M = 80$. Compared to the discrete processor, the experimental errors have more significant influence on the processing accuracy of the integrated processor. Note that we have not taken into account any additional processing errors resulting from factors that may degrade the cooperative operation of different wavelength channels, as discussed earlier. Considering these factors could lead to even higher errors.

To reduce the errors induced by imperfect response of experimental components, employing advanced mode-locking approaches [1] to reduce the noise of microcombs could be beneficial for both discrete and integrated processors. For integrated processors, the chirp of silicon EOM can be mitigated by using push-pull configurations as well as p-n depletion mode structure [47], and proper methods to calibrate the bias point [43]. The shaping errors of integrated spectral shapers can be alleviated via calibration procedures and gradient-descent control [43]. Integrated delay elements introduce additional loss especially when using a waveguide with high propagation loss, and adiabatic Euler bends can be employed to achieve low-loss and low-crosstalk waveguide bends [48]. The use of a wavelength-addressable serial integration scheme can also enable large-scale integration [50]. On the other hand, for discrete processors, there is still room for improving the processing accuracy. Errors of delay elements induced by higher-order dispersion can be reduced by using programmable phase characteristics of optical spectral shapers (OSSs), and the shaping errors can be minimized through employing feedback control [44]. These results have significant implications for all microwave photonics devices [51-97] based on microcomb sources [98-122] and other integrated nonlinear classical and quantum photonic chips [123-190].

## V. CONCLUSION

In summary, we provide a comparative study of the performance between microcomb-based RF photonic transversal signal processors implemented by discrete and integrated components. We first compare the performance of state-of-the-art processors, especially the processing accuracy. Next, analysis of multiple factors that contribute to the performance differences is conducted, including tap number and imperfect response of experimental components. Finally, we discuss the potential for future improvement. Our results show that although current integrated processors are attractive in providing significantly reduced system size, power consumption, and cost, their processing accuracy is not as high as the discrete processors. In addition, there is still room for improvement for both the discrete and integrated processors. The results offer valuable insights for microcomb-based RF photonic transversal signal processors with high reconfigurability for diverse applications.

**Conflict of Interest**

The authors declare that there is no conflict of interest for this paper.